\documentclass[10pt,superscriptaddress,twocolumn,amsmath,amssymb,aps,prb,reprint]{revtex4-1}

\usepackage{mathrsfs}
\usepackage{graphicx}
\usepackage{dcolumn}
\usepackage{bm}
\usepackage{color}

\usepackage{multirow}
\usepackage[colorlinks,bookmarks=false,citecolor=blue,linkcolor=red,urlcolor=blue]{hyperref}
\usepackage{ulem}

\renewcommand{\Im}{\operatorname{Im}}

\newcommand{\be}{\begin{equation}}
\newcommand{\ee}{\end{equation}}
\newcommand{\bea}{\begin{eqnarray}}
\newcommand{\eea}{\end{eqnarray}}

\definecolor{purple}{RGB}{128,0,128}

\begin{document}
\title{Pseudogap Behavior in the Local Spinon Spectrum of Power-Law Diverging Multichannel Kondo Model}
\author{Zuodong Yu}
\affiliation{School of Information and Electronic Engineering, Zhejiang Gongshang University, Hangzhou 310018, China}
\affiliation{National Laboratory of Solid State Microstructure, Department of Physics, Nanjing University, Nanjing 210093, China}
\author{Danqing Hu}
\affiliation{Department of Physics and Chongqing Key Laboratory for Strongly Coupled Physics, Chongqing University, Chongqing 401331, China}
\author{Jiangfan Wang}
\affiliation{School of Physics, Hangzhou Normal University, Hangzhou, Zhejiang 311121, China}
\author{Xinloong Han}
\email{hanxinloong@gmail.com}
\affiliation{Kavli Institute for Theoretical Sciences, University of Chinese Academy of Sciences, Beijing 100190, China}

\date{\today}
\begin{abstract}

Motivated by the emergence of higher-order van Hove singularities (VHS) with power-law divergent density of states (DOS) ($\rho_c(\omega)=\rho_0/|\omega|^{r}$, $0<r<1$) in materials, we investigate a multichannel Kondo model involving conduction electrons near the higher-order van Hove filling. This model considers $M$ channel and $N$ spin degrees of freedom. Employing a renormalization group analysis and dynamical large-$N$ approach, our results reveal a crossover from a non-Fermi liquid to pseudogap behavior in the spectral properties of the local impurity at the overscreened fixed point. We precisely determine the conditions under which the crossover occurs, either by tuning the exponent $r$ or the ratio $\kappa=M/N$ to a critical value. This pseudogap phase of spinon exhibits distinct physical properties that could have an impact on the properties of real systems. The results of this study provide novel insights into the non-Fermi liquid and pseudogap behaviors observed in strongly correlated systems and offer a playground to study the interplay between higher-order van Hove singularities and multichannel Kondo physics.

\end{abstract}
 \maketitle

\section{Introduction}
Understanding exotic phenomena in strongly correlated electronic systems is a central problem in condensed matter physics. Among these phenomena, the Kondo effect is of particular interest, describing the screening of a single magnetic impurity by itinerant electrons, resulting in the formation of a many-body singlet state below a characteristic Kondo temperature $T_{K}$ in metals \cite{hewson1997kondo}.
When the magnetic impurity is screened by multiple conduction channels symmetrically, intriguing physics emerges, including non-Fermi liquid (NFL) behaviors \cite{nozieres1980kondo,ludwig1991exact} and fractionalized quasiparticles \cite{emery1992mapping,lopes2020anyons,komijani2020isolating}. These exotic effects may have relevance for real heavy fermion materials \cite{cox1987quadrupolar,onimaru2016exotic}.
In certain limits, such as the large-$N$ limit, the multichannel Kondo model becomes exactly solvable, making it an ideal playground for studying strong electron correlation effects \cite{cox1993spin,parcollet1997transition,parcollet1998overscreened}. Unlike its single-channel counterpart, whose low-temperature properties can be described using Fermi liquid theory around a strong-coupling fixed point \cite{nozieres1974fermi-liquid}, the multichannel Kondo model exhibits a stable intermediate overscreening fixed point and a non-Fermi liquid ground state. Analytical studies based on conformal field theory have shed light on this intriguing behavior\cite{parcollet1998overscreened}.

In strongly correlated systems, two intriguing and extensively studied novel states are the non-Fermi liquid (NFL) metallic state and the pseudogap phase. The former is characterized by anomalous electrical transport and thermodynamic properties, exemplified by the $T$-linear resistivity observed in cuprates \cite{hill2001breakdown,cooper2009anomalous,yuan2022scaling}, iron-based superconductors \cite{doiron2009correlation,Dai2013Hidden}, and heavy-fermion systems \cite{custers2003break,gegenwart2008quantum,si2010heavy,badoux2016change,shen2020strange}. Despite several decades of theoretical studies \cite{sachdev2010strange,varma2020colloquium,phillips2022stranger}, its microscopic origin remains elusive.
In contrast, the pseudogap phase, associated with superconducting pairing, has been extensively investigated in cuprate superconductors \cite{timusk1999pseudogap,lawler2010intra,badoux2016change,zhao2017global,varma1989phenomenology,varma1999pseudogap,kyang2006_YRZ,Rice_2012}. A notable distinction between these two states lies in the low-energy behavior of their self-energies. The NFL state typically exhibits a sublinear power-law vanishing behavior, characterized by $\Sigma_{NFL}(\omega)\propto \omega^{\alpha_1}$, with a parameter $\alpha_1\in (0,1)$. On the other hand, the pseudogap phase features a diverging self-energy, generally following $\Sigma_{PG}(\omega)\propto \omega^{\alpha_2}$, where $\alpha_2\in\left[-1,0\right)$.
For instance, the self-energy of the pseudogap phase takes the form $\Sigma(\omega,{\bf k}) \propto [\omega+\varepsilon^{\prime}_{\bf k}]^{-1}$ in the doped RVB spin liquid with a kinetic energy $\varepsilon_{\bf k}=-2t (\cos(k_x)+\cos(k_y))$ involving nearest-neighbor hopping $t$ \cite{kyang2006_YRZ,Rice_2012}, as well as in the Hatsugai–Kohmoto model with $\varepsilon^{\prime}_{\bf k}=\varepsilon_{\bf k}+U/2$, where $\varepsilon_{\bf k}$ represents the bare kinetic energy, and $U$ denotes the long-range interaction strength \cite{phillips2020exact}.
Despite the intensive research on both the NFL and pseudogap phases, a simple model capable of accommodating both phenomena and capturing their transitions is still lacking.

The electron density of states (DOS) $\rho_c(\omega)=\rho_0|\omega|^{-r}$ (where $|\omega|\leq D$, and $D$ is the bandwidth) of the bath plays a crucial role in determining the low-temperature behaviors of the Kondo model. In the framework of the renormalization group (RG), different forms of DOS lead to distinct fixed points along the RG trajectories. For the case of a flat DOS ($r=0$), the RG analysis reveals an unstable local moment (LM) fixed point and a stable overscreened (OS) fixed point when the channel number exceeds $2S$ ($S$ is the impurity spin size) \cite{parcollet1998overscreened}. On the other hand, for the pseudogap DOS ($-1<r<0$), the RG flow diagrams exhibit more intricate structures: a LM fixed point at weak coupling and an OS phase at intermediate coupling \cite{vojta2001multichannel}.
However, investigations of the single-channel Kondo model with $S=1/2$ and single impurity Anderson model with a diverging DOS indicate the presence of a stable ferromagnetic coupling fixed point and an antiferromagnetic strong-coupling fixed point with a power-law dependence of the coupling strength on the Kondo temperature $T_K$ \cite{mitchell2013quantum,shankar2023}.

\begin{figure}
	\includegraphics[width=0.45\textwidth]{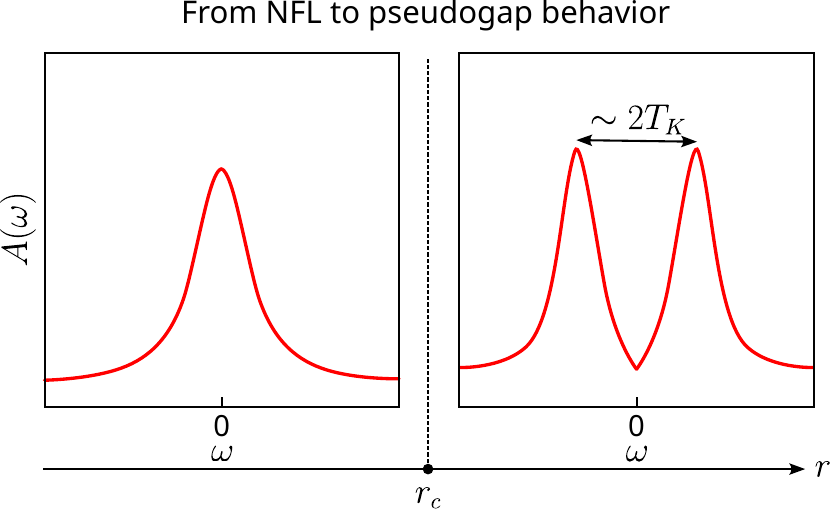}
	\caption{(color online) An illustration of the crossover from the non-Fermi liquid to the pseudogap phase by tuning the exponent $r$ of the electron density of states $\rho_c(\omega)\propto |\omega|^{-r}$ in the multichannel Kondo model with higher-oder VHS. $T_K$ is the Kondo temperature in the problem.}
	\label{fig:Fig1}
\end{figure}

In this study, we conduct a comprehensive investigation of the multichannel Kondo model with SU($N$) spin and SU($M$) channel symmetry, taking into account a power-law diverging density of states of conduction electrons in the large-N limit. The analytical results are obtained through RG analysis and scaling relations, further substantiated by numerical confirmation through dynamical large-N calculations. This allow an in-depth exploration of the dynamical properties of the model in proximity to the fixed point. Significantly, our findings reveal a crossover from the NFL phase to a pseudogap phase in the spectral function of the local impurity by tuning the exponent $r$ of the conduction electron DOS, as sketched in Fig.~\ref{fig:Fig1}. This pseudogap phase exhibits distinct resistivity and susceptibility behaviors compared to cases with constant or pseudogap DOS.

The structure of the paper is organized as follows: In Sec. II we introduce the large-N version of the power-law Kondo model and present the fermionic representation of $SU(N)$ spin. Section III is dedicated to RG analysis using a small $r$ expansion to derive the phase diagram of the model. We explore the dynamical properties of the fixed point in the antiferromagnetic (AFM) side, employing the scaling relation in the large-N limit in Section IV.  The results are further substantiated through numerical large-N calculations in Section V. To conclude, we encapsulate the paper with a discussion of $1/N$ corrections, the realization of the model in materials, and prospects for further studies in Section VI.

\section{Model}
We begin with the following large-$N$ Hamiltonian for the multichannel Kondo model\cite{parcollet1997transition,parcollet1998overscreened},
\bea
\mathcal{H}_{MK}=\sum_{\bf{k}}c^{\dagger}_{\alpha\sigma,{\bf k}}(\epsilon_{{\bf k}}-\mu) c_{\alpha\sigma,{\bf k}}+\frac{J_K}{N}\sum_{\alpha}\mathbf{S}\cdot\mathbf{s}_{\alpha},
\eea
where the symmetry of local spin is extended from $SU(2)$ to $SU(N)$. We choose anti-symmetric representation of $SU(N)$ which in terms of pseudofermions (spion) operators $f_{\sigma}$ as  $S_{\sigma\sigma^{\prime}}=f^{\dagger}_{\sigma}f_{\sigma^{\prime}}-Q/N\delta_{\sigma\sigma^{\prime}}$ where $\sigma=1,\dots,N$ the spin flavor and $Q=\sum_{\sigma}f^{\dagger}_{\sigma}f_{\sigma}$. Here we choose particle-hole symmetric case where $Q=N/2$, enforced through a Lagrange multiplier $\lambda$. Accordingly, the channel is also extended to $SU(M)$, where $\alpha=1,\dots,M$ is the indices of charge channel. In order to get meaningful result, the Kondo interaction, which corresponding to the coupling strength between the local magnetic impurity and conducting electrons, is scaled to $J_K/N$. This allows us to express the Kondo term as\cite{parcollet1998overscreened},
\bea
\frac{J_K}{N}\sum_{\alpha}\mathbf{S}\cdot\mathbf{s}_{\alpha}=\frac{J_K}{N}\sum_{\sigma\sigma^{\prime}} (f^\dagger_\sigma f_{\sigma^{\prime}}-\delta_{\sigma\sigma^{\prime}}/2) c^{\dagger}_{\alpha\sigma^{\prime},{\bf k}} c_{\alpha\sigma,{\bf k}^{\prime}}.
\label{eq:kondo}
\eea
The $c_{\alpha\sigma,\mathbf{k}}$ represents the electron annihilation with the dispersion $\epsilon_{\mathbf{k}}$ and the chemical potential $\mu$. This large-N model possess the $SU(N)\times SU(M)$ symmetry. In the calculation, we fixed the ratio $\kappa=M/N$ when take large-N limit. A key feature of our investigation involves considering a power-law divergent density of states (DOS), characterized by $\rho_{c}(\omega)=\rho_0/|\omega|^{r}$, where $0<r<1$ is a positive number. The value $r=1/3$ corresponds to the case of biased bilayer graphene \cite{PhysRevB.95.035137}, while $r=1/4$ pertains to kagome metals AV$_3$Sb$_5$ (A=K, Rb, Cs) and `magic' twisted bilayer graphene due to the higher-order van Hove singularities (VHS) \cite{kang2022twofold,hu2022rich,PhysRevB.107.184504,shankar2023}.

We introduce Green's functions $G_{c}(\tau,{\bf k})=-\langle \mathcal{T}_{\tau} c_{\alpha\sigma}(\tau)c^{\dagger}_{\alpha\sigma}(0)\rangle$ and $G_f(\tau)=-\langle \mathcal{T}_{\tau} f_{\alpha}(\tau)f^{\dagger}_{\alpha}(0)\rangle$, which describe the conduction electrons and spinons, respectively. The bare Green's function are
\bea
G_{c}(i\omega_n)=\frac{1}{i\omega_n-E_{\bf k}}, G_{f,0}(i\omega_n)=\frac{1}{i\omega_n-\lambda}
\eea
with $E_{\bf k}=\epsilon_{\bf k}-\mu$ and fermionic Matsubara frequency $i\omega_{n}$. Our analysis focuses on the AFM Kondo coupling in the large-N limit with a fixed value of $\kappa$. We perform a renormalization group (RG) analysis and then present the dynamical and transport properties in the vicinity of the crossover region, employing the dynamical large-N approach.

\begin{figure}[ht]
	\includegraphics[width=0.45\textwidth]{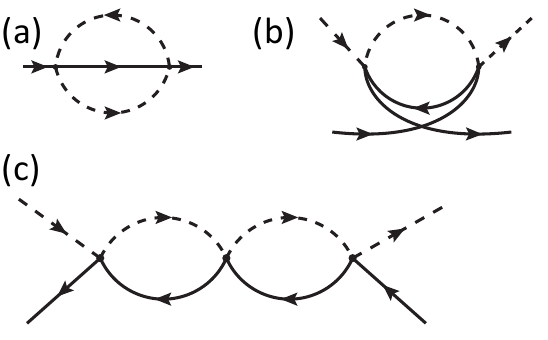}
	\caption{(color online) Feynman diagrams in a large-$N$ limit where the black solid or dashed line represents the propagator of fermionic spinons or conduction electrons. (a) Correction to $Z_f$. (b) Correction to $Z_J$ to the first order in $J_K$. (c) Correction to $Z_J$ to the second order in $J_K$.}
	\label{fig:Fig2}
\end{figure}
\section{RG analysis by small $r$ expansion}\label{RG}
In this section we perform a renormalization group analysis based on dimensional regularization with minimal subtraction of poles\cite{zinn2021quantum,zhu2002critical}. Defining the renormalization field $f_{\sigma,\Lambda}$ and dimensionless coupling constant $J_{K,\Lambda}$ at an energy cutoff $\Lambda$ running from the bandwidth $D$ to 0, as $f_{\sigma}=Z_f^{1/2} f_{\sigma,\Lambda}$ and $J_K=Z_J Z_f^{-1} J_{K,\Lambda} \Lambda^{r}$, respectively, where $Z_f$ and $Z_J$ represent the renormalization factors for $f_{\sigma,\Lambda}$ and $J_{K,\Lambda}$. Within this calculation, the Hamiltonian will also contain counterterms which remove the singular part during RG calculations, and this counterterm Hamiltonian\cite{zhu2002critical} is
\bea
H_{CT}=\sum_{\sigma}(Z_f-1)\lambda f_{\sigma}^{\dagger} f_{\sigma}+(Z_J-1)\frac{J_{K,\Lambda}\Lambda^{r}}{N}\sum_{\alpha}{\bf S}\cdot {\bf s}_{\alpha},
\eea
We firstly perform the renormalization of the Green's function of $f$, it is given by
\bea
G_f^{-1}(\omega+\lambda)=\omega-\Sigma_f(\omega+\lambda)=Z_f G_{f,0}^{-1}(\omega+\lambda).
\eea
In the large-$N$ limit, the self-energy $\Sigma_{f}(\omega)$ is contributed by the diagram Fig. \ref{fig:Fig2} (a). Hence the self-energy at the finite temperature is,
\bea
&&\Sigma_f(i\omega_n)=-\frac{J^2_{K,\Lambda}\Lambda^{2 r}}{N^2}\frac{T^2}{V^2}\sum_{\sigma,\alpha}\sum_{{\bf k},{\bf k}^{\prime}}\sum_{i\omega_1,i\omega_2}G_{f}(i\omega_1)\nonumber \\
&&\times G_c(i\omega_2,{\bf k})G_c(i\omega_f+i\omega_2-i\omega_1,{\bf k}^{\prime})\nonumber \\
&&=J_{K,\Lambda}^2\kappa \Lambda^{2 r}\sum_{{\bf k},{\bf k}^{\prime}}\frac{f(E_{\bf k}) f(-E_{{\bf k}^\prime})}{i\omega_n-\lambda+E_{\bf k}-E_{{\bf k}^\prime}},
\eea
where $f(\omega)=1/(e^{\omega/T}+1)$ is the Fermi-Dirac distribution function. $\omega_{n,1,2}$ is the Fermionic Matasubara frequency. At the zero temperature we have
\bea
&&\Sigma_f(\omega+\lambda)=-\int_{\mathcal{R}} d\omega_1 d \omega_2 \frac{J_{K,\Lambda}^2\Lambda^{2 r} \kappa \rho_0^2}{|\omega_1|^{\epsilon^{\prime}}|\omega_2|^{\epsilon^{\prime}}(\omega+\omega_1-\omega_2)}\nonumber \\
&&=-J_{K,\Lambda}^2 \kappa \rho_0^2\omega\Gamma(2r-1)(\frac{\Lambda}{\omega})^{2r}\nonumber \\
&&\simeq \frac{J_{K,\Lambda}^2  \rho_0^2\kappa }{2r}\omega+J_{K,\Lambda}^2 \kappa \rho_0^2\omega \ln(\frac{\Lambda}{\omega}),
\eea
where $\int_{\mathcal{R}} d\omega_1\omega_2=\int_{-\infty}^0d\omega_1\int_0^{+\infty}d\omega_2$. $\Gamma(x)$ is the usual Euler Gamma function. And we use the following integral,
\bea
\int_0^1 d x(1-x)^{\epsilon_1-1} x^{\epsilon_2-1}=\frac{\Gamma(\epsilon_1)\Gamma(\epsilon_2)}{\Gamma(\epsilon_1+\epsilon_2)}.
\eea
By demanding the poles of the self-energy can be cancelled, the $Z_f=1-\Sigma_f(\omega+\lambda)/\omega$ becomes,
\bea
Z_{f}=1-\frac{\kappa\rho_0^2 J_{K,\Lambda}^2}{2 r}.
\eea
The first order correction to the interaction vertex $\Gamma_J\equiv J_{K,\Lambda} \Lambda^{r}\gamma_J$ in the large-$N$ limit is described by Fig. \ref{fig:Fig2}(b), which reads
\bea
\gamma_J^{(1)}(i\omega_b)=J_{K,\Lambda} \Lambda^r\Pi_{ZS}(i\omega_b),
\eea
where $i\omega_b$ is the Bosonic Matasubara frequency. The particle-hole bubble $\Pi_{ZS}(i\omega_b)$ is defined as following
\bea
\Pi_{ZS}(i\omega_b)&&=\frac{T}{V}\sum_{i\omega_1,{\bf k}}G_{f}^0(i\omega_1)G_c(i\omega_1-i\omega_b,{\bf k}) \nonumber \\
&&=\frac{1}{V}\sum_{\bf k} \frac{f(E_{\bf k})}{i\omega_b+E_{\bf k}-\lambda}.
\eea
Then we have
\bea
\gamma_J^{(1)}(\omega+\lambda)&&=J_{K}\rho_0\Gamma(r)\Gamma(1-r)(\frac{\Lambda}{\omega})^r\nonumber \\
&&\simeq \frac{J_{K}\rho_0}{r}+ J_{K}\rho_0\ln(\frac{\Lambda}{\omega}),
\eea
and the second order correction in $J_{K,\Lambda}$ to the interaction vertex is illustrated in Fig. \ref{fig:Fig2}(c). It can be straightforwardly calculated as 
\bea
\gamma^{(2)}_J(\omega+\lambda)=-J_{K,\Lambda}^2\Pi_{ZS}^2(\omega+\lambda)=-\frac{J_{K,\Lambda}^2\rho_0^2}{r^2}(\frac{\Lambda}{\omega})^{2r}.\nonumber \\
\eea
To cancel out the poles in $\gamma_J^{(1)}$ and $\gamma_J^{(2)}$, $Z_J$ becomes
\bea
Z_J=1-\frac{J_{K}\rho_0}{r}+\frac{J_{K,\Lambda}^2\rho_0^2}{r^2}.
\eea
In the framework of renormalization group procedure, the bare vertex interaction $J_K$ remains the same, and the beta function for $J_{K,\Lambda}$ is defined as $\beta(J)= \partial J_{K,\Lambda}/\partial \ln(D/\Lambda)$ which is determined by\cite{zhu2002critical}
\bea
\frac{d}{d\ln(D/\Lambda)}\big(Z_J Z_f^{-1} J_{K,\Lambda} \Lambda^{r}\big)=0,
\eea
which leads to
\bea
\beta(J)=\frac{r}{1/J_{K,\Lambda}+\frac{\partial\ln(Z_J Z_f^{-1})}{\partial J_{K,\Lambda}}}.
\eea
\begin{figure}[ht]
	\includegraphics[width=0.45\textwidth]{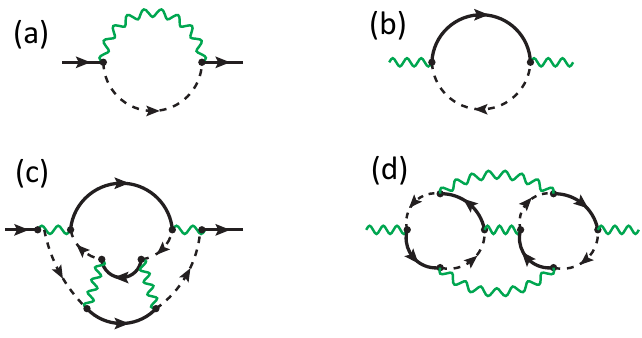}
	\caption{(color online) The leading Feynman diagrams in large-$N$ limit for the self-energy of (a) $f$ and  (b) $B$ field respectively. (c), (d): one of the $1/N$ correction diagrams for $f$ and $B$.  The green wavy line denotes the propagator of $B$-field.}
	\label{fig:Fig3}
\end{figure}
By expanding the above equation to the third order in $J_{K,\Lambda}$, we can obtain
\bea
\frac{\partial \bar{J}_K}{\partial \ln(D/\Lambda)}=r \bar{J}_K+\bar{J}_K^2-\kappa \bar{J}_K^3,
\eea
where the dimensionless coupling constant $\bar{J}_K=J_{K}\rho_0$. Notably, there exists a stable fixed point at $\bar{J}_{OS}=(1+\sqrt{1+4\kappa r})/2\kappa$, representing the overscreened Kondo phase, and an unstable trivial fixed point $\bar{J}_{LM}=0$ indicating a local moment phase. Moreover, our RG analysis reveals the existence of a stable ferromagnetic coupling fixed point with $\bar{J}_{FM}=(1-\sqrt{1+4\kappa r})/2\kappa$. In this work, we focus on the overscreened fixed point $J_{OS}$, which governs the low-energy physics in the antiferromagnetic coupling regime.

\section{Large-$N$ limit with AFM coupling}
To explore the Kondo overscreened phase with AFM coupling both analytically and numerically in the large-$N$ limit, we introduce charged bosonic fields in each channel $B_{\alpha}(\tau)$ which conjugates to $\sum_{\sigma,{\bf k}} f^{\dagger}_{\sigma}(\tau) c_{\alpha \sigma,{\bf k}}(\tau)/\sqrt{N}$, enabling the decoupling of the Kondo interaction Eq.~(\ref{eq:kondo})\cite{parcollet1998overscreened,zhu2004quantum}:
\bea
\sum_{\alpha}[\bar{B}_{\alpha}\frac{1}{J_{K}}B_{\alpha}+\frac{1}{\sqrt{N}}\sum_{\sigma}(f_{\sigma}^{\dagger}c_{\sigma,\alpha}B_{\alpha}+\bar{B}_{\alpha}c_{\sigma,\alpha}^{\dagger}f_{\sigma})].
\eea
We define the Greens' function $G_B(\tau)$ of the bosonic fields $B_{\alpha}$ as $G_B(\tau)=\sum_{\alpha}\langle \mathcal{T}_{\tau} B_{\alpha}(\tau) B^{\dagger}_{\alpha}(0)\rangle/M$. 
In the large-$N$ limit, the saddle point equations are given by\cite{yu2020quantum}
\bea\label{Eq:Selfenergies1}
&&G_{f}^{-1}(i\omega_n)=i\omega_n-\lambda-\Sigma_f(i\omega_n),\\
&&G_B^{-1}(i\Omega_n)=G^{-1}_{B,0}-\Pi(i\Omega_n),
\eea
with $G_{B,0}^{-1}(i\Omega_n)=-1/J_{K}$. The self-energies of $f_{\sigma}$ and $B_{\alpha}$ are given by
\bea
&&\Sigma_f(i\omega_n)=\frac{T}{N}\sum_{\alpha}\sum_{i\omega_m}G_0(i\omega_m) G_B(i\omega_n-i\omega_m), \label{Eq:Selfenergies2_1}\\
&&\Pi(i\Omega_n)=-\frac{T}{N}\sum_{\sigma}\sum_{i\omega_m}G_0(i\omega_m) G_f(i\omega_m-i\Omega_n),\label{Eq:Selfenergies2}
\eea
which are represented by the Feynman diagrams in Fig.~\ref{fig:Fig3}(a,b). $i\omega_{m,n}$ is the fermionic Matsubara frequency and $i\Omega_n$ is the bosonic Matsubara frequency. And the Green's function for the conduction electron $G_0(i\omega_n)$ is defined as 
\bea\label{Eq:SpectralR}
G_0(i\omega_n)\equiv\frac{1}{V}\sum_{{\bf k}}\frac{1}{i\omega_n-E_{\bf k}}=\int d\omega^{\prime}\frac{\rho_c(\omega^{\prime})}{i\omega_n-\omega^{\prime}}.
\eea
Before employing the numerical dynamical large-N calculation for the above saddle point equations Eq. (\ref{Eq:Selfenergies1}-\ref{Eq:Selfenergies2}) to obtain the Green's functions $G_f$ and $G_B$, we first consider the case in low-energy and zero temperature limit. It is more convenient to rewrite the saddle point equations in the imaginary time as
\bea\label{Eq:tausigma}
\Sigma_f(\tau)=-\kappa G_{0}(\tau) G_B(\tau), \Sigma_B(\tau)=G_{0}(\tau) G_f(-\tau).
\eea
In the low temperature regime $T_K^{-1}\ll\tau\ll T^{-1} \rightarrow \infty$ where $T_K$ is the Kondo temperature acting as short-time cutoff of the problem, it is expected that self-energy governs the saddle point equations as\cite{parcollet1998overscreened}
\bea\label{Eq:scalingform}
G_{f}(i\omega_n)\Sigma_f(i\omega_n)=-1, \Pi(i\Omega_n)G_{B}(i\Omega_n)=-1.
\eea
By solving spectral representation Eq. (\ref{Eq:SpectralR}), the local Green's function for conduction electrons is $G_0(i\omega_n)=C\rho_0|\omega_n|^{-r}sgn(\omega_n)$ with constant $C$. In the low energy and zero temperature limit, we assume the following scaling forms,
\bea
&&G_f(i\omega_n)=iA_1 |\omega_n|^{\Delta_1-1} sgn(\omega_n),\label{Eqs:scalingGf}\\
&&G_B(i\Omega_n)=A_2|\Omega_n|^{\Delta_2-1},\label{Eqs:scalingGB}
\eea
where $A_{1,2}$ are prefactors and $\Delta_{1/2}>0$ are the scaling exponents to be determined. The retarted Green's function can be obtained by analytical continuation as $G^R(\omega)=G(\omega+i\eta)$ with the following results,
\bea
G_{f}^R(\omega)=-\frac{A_1 e^{-i\frac{\Delta_1}{2}\pi}}{(\omega+i\eta)^{1-\Delta_1}}, G_B^{R}(\omega)=i\frac{A_2e^{-i\frac{\Delta_2}{2}\pi}}{(\omega+i\eta)^{1-\Delta_2}},
\eea
with a positive infinitesimal value $\eta$. Thus in the low-energy and zero temperature limit, the spectral function for $f$ or $B$ field $A_{f,B}(\omega)=-\Im G_{f,B}^R(\omega)/\pi$  is 
\bea
A_{f}(\omega)=-\frac{A_1\sin(\frac{\Delta_1}{2}\pi)}{\pi|\omega|^{1-\Delta_1}}, A_B(\omega)=-\frac{A_2\cos(\frac{\Delta_1}{2}\pi)sgn(\omega)}{\pi|\omega|^{1-\Delta_2}}, \nonumber \\
\eea
\begin{figure}[ht]
	\includegraphics[width=0.5\textwidth]{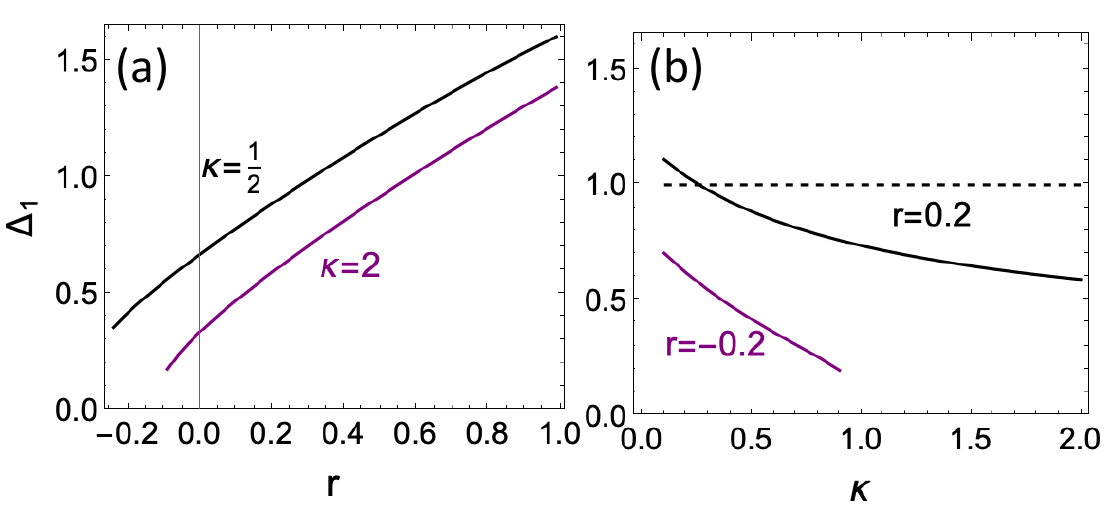}
	\caption{(color online) The scaling exponent $\Delta_1$ as functions of (a) $r$ and (b) $\kappa=M/N$ by solving Eqs (\ref{eq:exponents1}) and (\ref{eq:exponents2}).}
	\label{fig:Fig4}
\end{figure}
 After doing Fourier transformation for Eqs. (\ref{Eqs:scalingGf}) and (\ref{Eqs:scalingGB}), the imaginary time Green's functions can be obtained as
\bea
&&G_{f}(\tau)=\frac{A_1}{2\cos(\frac{\Delta_1}{2}\pi)\Gamma(1-\Delta_1)}\frac{sgn(\tau)}{|\tau|^{\Delta_1}},\label{Eq:tauGf}\\
&&G_{B}(\tau)=\frac{A_2}{2\sin(\frac{\Delta_2}{2}\pi)\Gamma(1-\Delta_2)}\frac{1}{|\tau|^{\Delta_2}},\label{Eq:tauGB}
\eea
and $G_0(\tau)\sim sgn(\tau)/|\tau|^{1-r}$. Here we use the following integrals,
\bea
&&\int d\tau e^{i\omega \tau} \frac{sgn(\tau)}{|\tau|^{\Delta}}=i2\frac{\Gamma(1-\Delta)\cos(\frac{\Delta}{2}\pi)}{|\omega|^{1-\Delta}}sgn(\omega),\\
&&\int d\tau e^{i\omega \tau} \frac{1}{|\tau|^{\Delta}}=2\frac{\Gamma(1-\Delta)\sin(\frac{\Delta}{2}\pi)}{|\omega|^{1-\Delta}}.
\eea
Putting back Eqs. (\ref{Eqs:scalingGf}, \ref{Eqs:scalingGB}, \ref{Eq:tauGf},\ref{Eq:tauGB}) into Eqs. (\ref{Eq:tausigma},\ref{Eq:scalingform}), we find that $\Delta_{1/2}$ must satisfy the following equations,
\bea\label{Eq:ScalingEq}
&&\Delta_1+\Delta_2=1+r, \label{eq:exponents1}\\
&& \kappa \Gamma(\Delta_1-1) \Gamma(1-\Delta_1)\cos^2(\frac{\Delta_1\pi}{2})= \nonumber \\
&&\Gamma(r-\Delta_1)\Gamma(\Delta_1-r)\cos^2(\frac{r-\Delta_1}{2}\pi).\label{eq:exponents2}
\eea

At $r=0$, there exists only one solution\cite{parcollet1998overscreened} $\Delta_1=1/(1+\kappa)$ and $\Delta_2=\kappa/(1+\kappa)$, coinciding with the fact that the oversreened phase is the only fixed point for the flat DOS conduction bath. We now fixed $\kappa=1/2$ and tune $r$ slightly away from zero, say $r=0.1$, the exponent $\Delta_1$ extracted from Eqs (\ref{eq:exponents1}) and (\ref{eq:exponents2}) is approximately $0.767$, which is larger than $2/3$. As we reach a critical value of the diverging DOS at $r_c\approx 0.314$ as shown in Fig. \ref{fig:Fig4}, the crossover from NFL to pseudogap phase emerges.  Importantly, our analysis reveals that the strongly divergent DOS of conduction electrons significantly enhances the corrections to the self-energy $\Sigma_f(\omega)$ of spinons, ultimately resulting in singular behavior in the pseudogap region. In Fig.~\ref{fig:Fig4}, we present the solution for the exponent $\Delta_1$ across the range of $r$ and $\kappa$. It is evident that the critical value of $r$ depends on $\kappa$ as well.

\begin{figure}[ht]
	\includegraphics[width=0.48\textwidth]{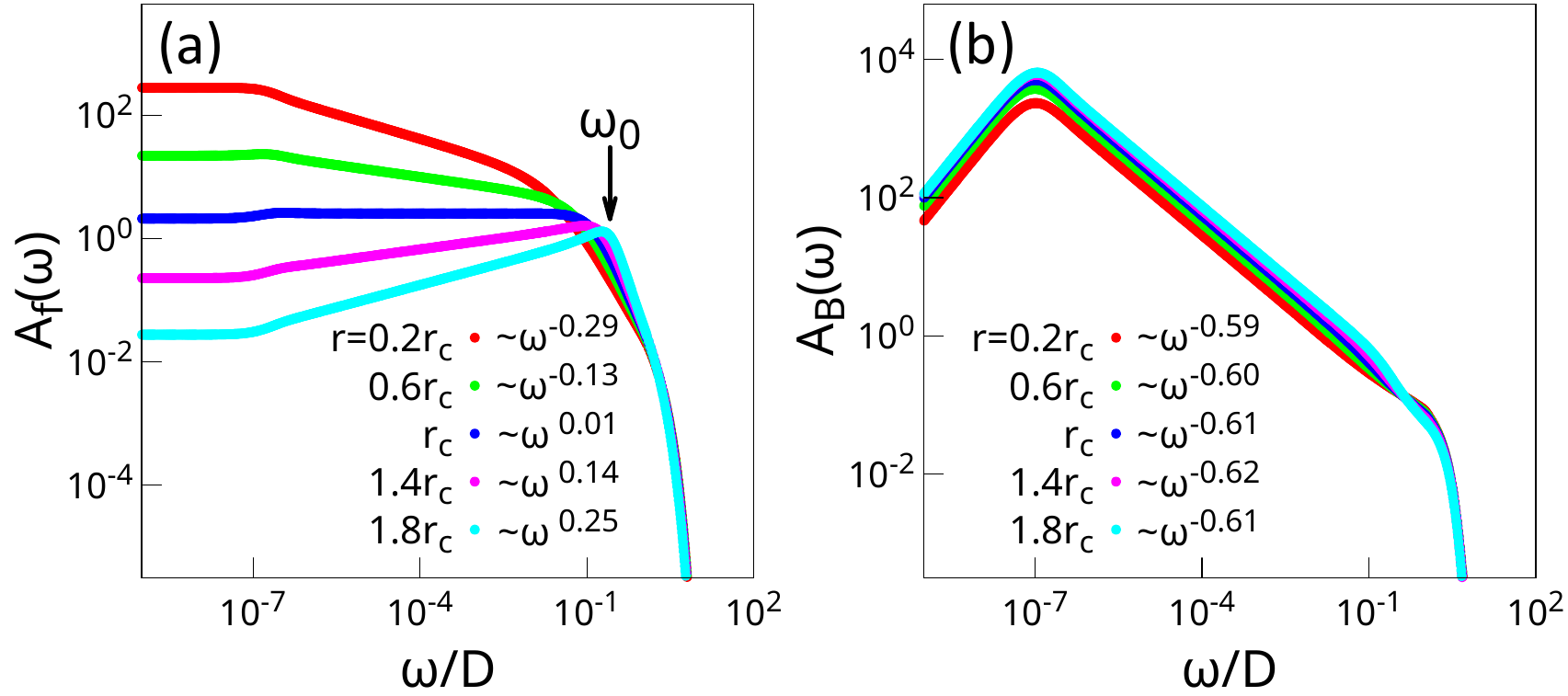}
	\caption{(color online) (a) The $r$ dependency of $A_f(\omega)$ with corresponding powerlaw exponent in scaling region. (b) the $r$ dependency of $A_B(\omega)$ with corresponding powerlaw exponent in scaling region. The temperature is set to be $10^{-7}D$.}
	\label{fig:Fig5}
\end{figure}

Now let's turn to discuss the observable in the multichannel Kondo model with power-law diverging DOS. One crucial observable is the resistivity arising from scattering of magnetic impurity. In the dilute limit, this contribution is given by\cite{cox1993spin,kim1997one}:
\bea\label{Eq:rho}
\rho(T)\sim -\left[\int d\omega \frac{\partial f(\omega)}{\partial \omega}\frac{ \rho_{c}(\omega)}{\mathrm{Im}[\mathcal{T}(\omega)]}\right]^{-1}.
\eea
Here, $\mathcal{T}(\omega)$ represents T-matrix for conduction electrons. In large-$N$ limit, the impurity contribution to conduction election is of $1/N$ order. Nevertheless, it has been demonstrated that $1/N$ corrections are irrelevant for power-law exponent of Green's functions, making them valid for finite $N$ as well\cite{cai2020dynamical,cox1993spin} (refer to the discussion in Section \ref{conclusion}). Thus in large-N, the T-matrix is given by the convolution of $G_{f}$ and $G_{B}$, which in imaginary time is defined as\cite{parcollet1998overscreened}
\bea\label{Eq:Tmat}
\mathcal{T}(\tau)=-G_{f}(\tau)G_{B}(-\tau).
\eea 

In the case of flat DOS ($r=0$), the resistivity $\rho(T)$ approaches a constant in the low-temperature limit\cite{parcollet1998overscreened,cox1993spin}, given by $\rho(T)=\rho(0)[1-\gamma (T/T_{K})^{\kappa/(\kappa+1)}]$ for $\kappa<1$, where $\gamma$ is a constant and $T_K$ is the Kondo scale in the considered problem. On the other hand, in the pseudogap case ($r<0$) at the scaling invariant oversreened fixed point, the resistivity diverges following a power-law behavior, $\rho(T)\propto (T_K/T)^{-2r}$, where the $-2r$ contribution arises from the pseudogap DOS and the scattering matrix $\mathcal{T}(\omega,T)$.

In the diverging DOS case ($r>0$), the resistivity exhibits distinct behaviors in comparison to the flat and pseudogap multichannel Kondo models.  It vanishes at zero temperature, following the form $\rho(T)\propto (T/T_K)^{2r}$, which deviates significantly from the Fermi liquid for $r\ll 1$. This nontrivial power-law behavior of $\rho(T)$ in the DPLMCK, which differ from the $\rho(T)\propto T^{1-r}$ in the absence of local impurity\cite{isobe2019supermetal}, signifies that the Kondo screening effect strongly modify the temperature behavior of resistivity.

\begin{figure}[ht]
	\includegraphics[width=0.48\textwidth]{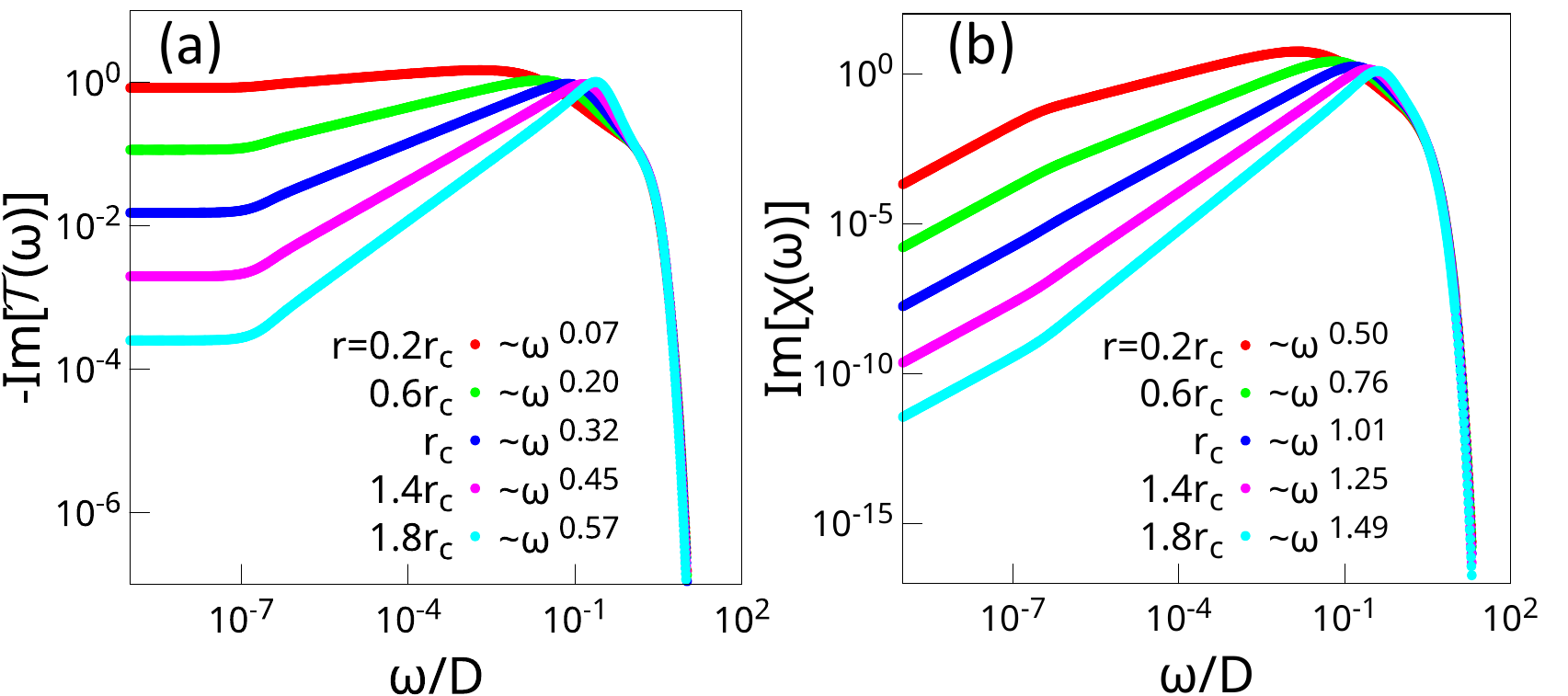}
	\caption{(color online) (a) The $r$ dependency of the imaginary part of T-matrix. with corresponding power law exponent in scaling region (b) the $r$ dependency of $\mathrm{Im}[\chi(\omega)]$ with corresponding power law exponent in scaling region. The temperature is set to be $10^{-7}D$.
	}
	\label{fig:Fig6}
\end{figure}

The dynamical susceptibility of the local impurity\cite{zhu2004quantum}
\bea\label{Eq:chi}
\chi(\tau)=-G_f(\tau) G_f(-\tau),
\eea
also exhibits anomalous scaling behavior. In the low-energy and zero-temperature limit, $\chi(\omega)\propto \omega^{1+\eta}$, where $\eta$ is a positive anomalous scaling exponent given by $\eta=2\Delta_1-2$ in the pseudogap phase when $r>r_c$. At the critical value $r_c$, $\eta$ vanishes, and the dynamic susceptibility becomes linear-in-$\omega$ at low energy, i.e., $\chi(\omega)\propto \omega$. This peculiar anomalous scaling behavior of the dynamic susceptibility can serve as a hallmark for diagnosing the occurrence of the crossover from NFL to the pseudogap phase in experiments. This constitutes one of the key points of this work.

\section{Numerical results}
To validate our analytical analysis and elucidate the crossover beyond the low-energy and temperature limits, we numerically solve the dynamical large-$N$ equations in real frequency space across a broad temperature range. The parameters are fixed at $\kappa=1/2$ and $J_{K}/D=0.8$ throughout the entire calculation unless explicitly stated otherwise. we employ a logarithmically dense frequency grid, and speed up convergence using a modified Broyden’s scheme\cite{zitko2009convergence}. Through analytical continuation, the self-energies in Eqs.~(\ref{Eq:Selfenergies2_1}, \ref{Eq:Selfenergies2}) become:
\begin{equation}\begin{split}
	\mathrm{Im}\Sigma_B^{R}(\nu)=&\frac{1}{\pi}\int_{x}\mathrm{Im}G_{f}^{R}(x)\mathrm{Im}G_{0}^{R}(x-\nu)[f(x)-f(x-\nu)],\\
	\mathrm{Im}\Sigma_{f}^{R}(\omega)=&\frac{-\kappa}{\pi}\int_{x}\mathrm{Im}G_{B}^{R}(x)\mathrm{Im}G_{0}^{R}(\omega-x)[b(x)+f(x-\omega)].
\end{split}\end{equation}
The real parts are given by Kramers-Kronig relation. Once the self-consistent solution is obtained, the local spin susceptibility and T-matrix define in Eqs.~(\ref{Eq:Tmat},\ref{Eq:chi}) is calculated as follows,
\begin{figure}[ht]
	\includegraphics[width=0.48\textwidth]{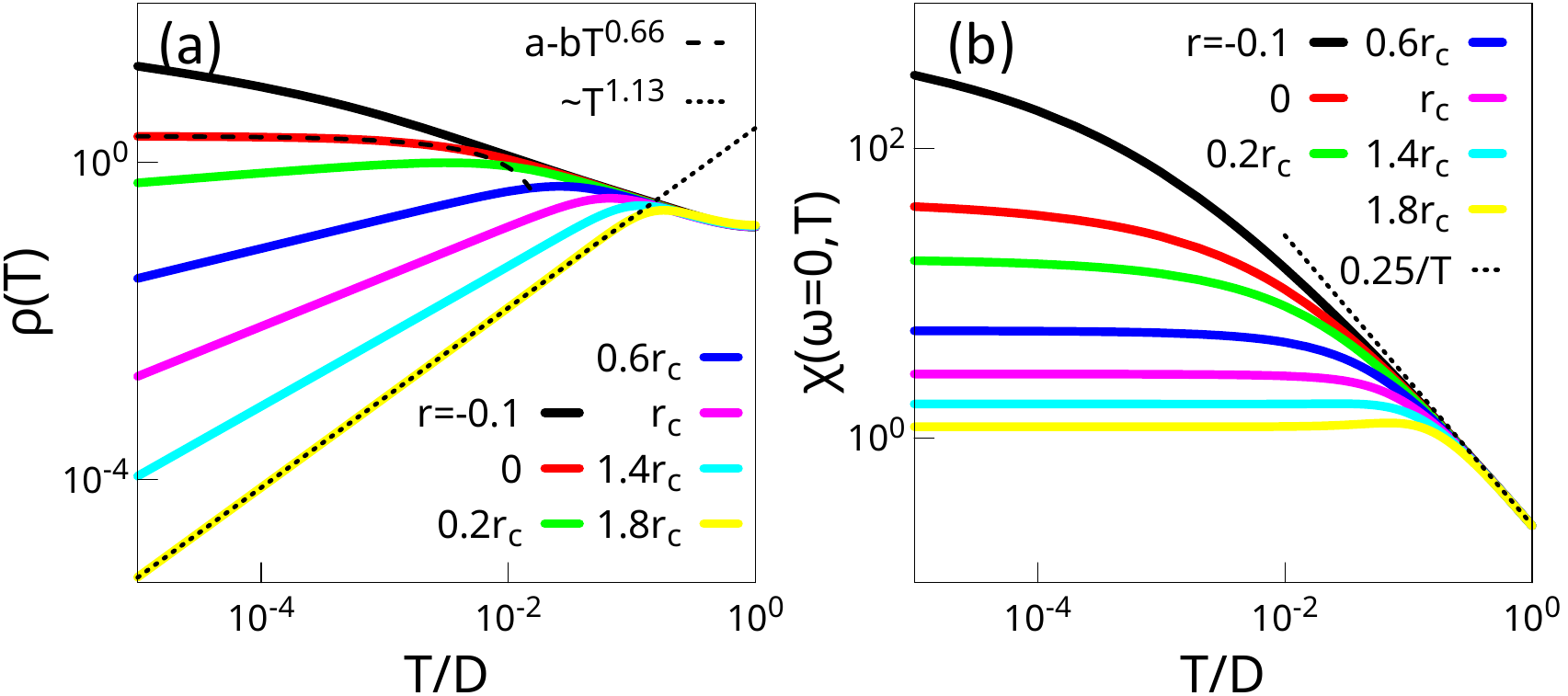}
	\caption{(color online) (a) The temperature dependency of resistivity for different $r$.  The dashed and dotted line are the fitting for low temperature. (b) The temperature dependency of static local susceptbility for different $r$.
	}
	\label{fig:Fig7}
\end{figure}

\begin{equation}\begin{split}\label{Eq:chi_Tmat}
	\mathrm{Im}\chi(\nu)=&\frac{1}{\pi}\int_{x}\mathrm{Im}G_{f}^{R}(x)\mathrm{Im}G_{f}^{R}(x-\nu)[f(x-\nu)-f(x)],\\
	\mathrm{Im}\mathcal{T}(\omega)=&\frac{-1}{\pi}\int_{x}\mathrm{Im}G_{f}^{R}(x)\mathrm{Im}G_{B}^{R}(x-\omega))[b(x-\omega)+f(x)].
\end{split}\end{equation}
The impurity entropy, defined as $s=S-S_0$ where $S$ is the total entropy and $S_0$ is the entropy in the absence of impurity, can be obtained using the Kadanoff-Baym formalism\cite{yu2020quantum,coleman2005sum} as
\begin{equation}\begin{split}\label{eq:S}
	s=&-\int \frac{d\omega}{\pi}\frac{db}{dT} \kappa \left[\mathrm{Im}\ln(-1/G_{B}^{R})+\mathrm{Im}\Sigma_{B}^{R}\mathrm{Re}G_{B}^{R} \right]\\
			 &+\frac{df}{dT} \left[\mathrm{Im}\ln(-1/G_{f}^{R})+\mathrm{Im}\Sigma_{f}^{R}\mathrm{Re}G_{f}^{R}-\kappa\mathrm{Re}\tilde{\Sigma}_{c}^{R}\mathrm{Im}G_{0}^{R} \right],
\end{split}\end{equation}
where $\tilde{\Sigma}_{c}^{R}(\omega)=\mathcal{T}(\omega)$ and $b(\omega,T)$ is the Boson-Einstein distribution function.  In Fig. \ref{fig:Fig5}(a), we present the spectral function of spinons $A_f(\omega,T)$ at different $r$ at relatively low temperature. As seen in the figure, the scaling behavior is evident for all value of $r$ considered here, consistent with our scaling analysis. Moreover, the pseudogap phase with positive scaling exponent begins at a critical value that agree well with the analytical $r_c$ obtained above. This manifests as a peak at finite energy $\omega_0$ as marked in Fig. \ref{fig:Fig5}(a). Interestingly, we observe that $\omega_0\sim T_K$ with $T_K$ as Eqs. \ref{eq:Tk} derived in Appendix~\ref{app} in the case of a power-law diverging DOS. The remarkable results suggests that $T_K$ is the only relevant energy scale in our model. The exponent extracted from the scaling region of $A_{f,B}(\omega)$ in Fig. \ref{fig:Fig5}(a,b) through numerical fitting also agree well with the solution of Eqs. (\ref{eq:exponents1}) and (\ref{eq:exponents2}).

\begin{figure}[ht]
	\includegraphics[width=0.48\textwidth]{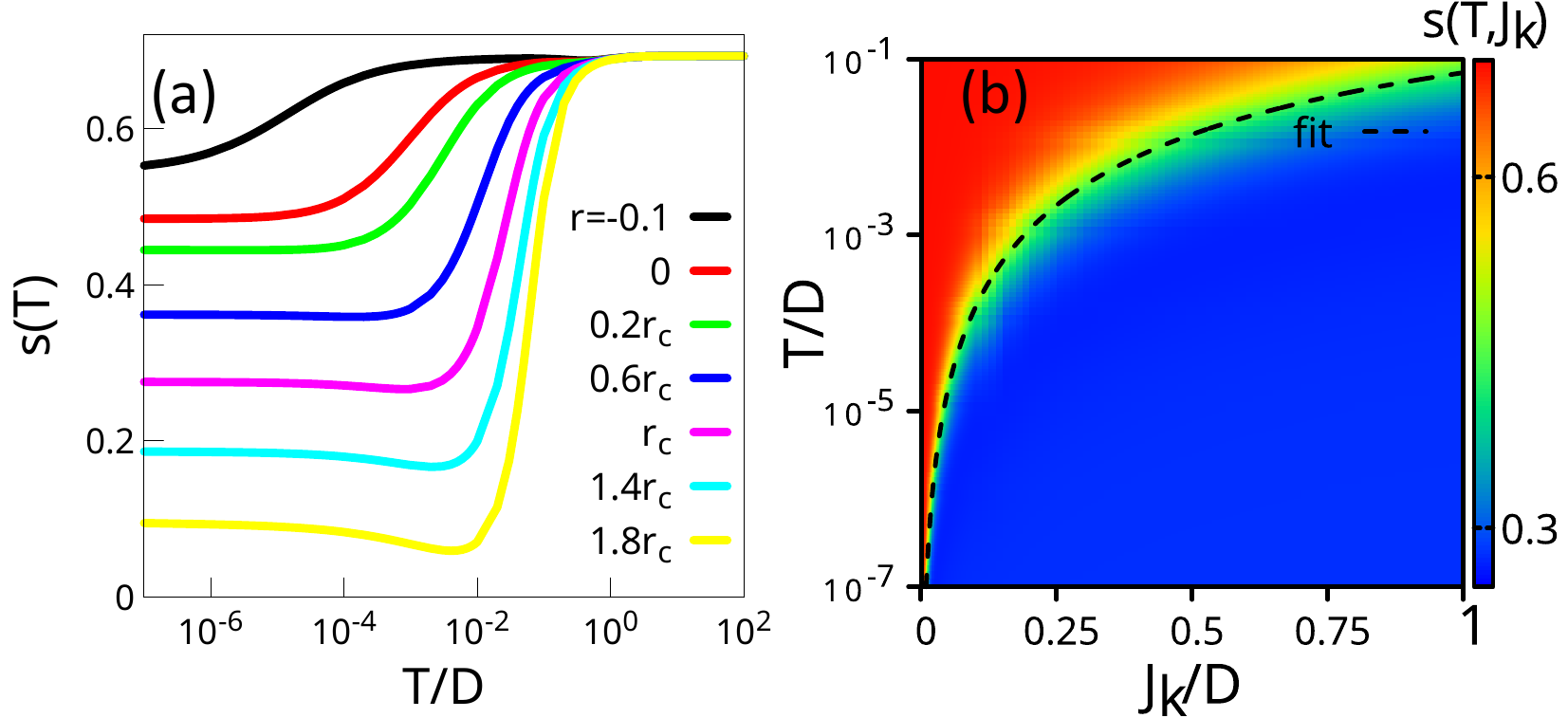}
	\caption{(a) The temperature dependency of impurity entropy for different $r$. (b) $s(T,J_{K})$ as function of temperature and Kondo interaction for $r=r_c$. The dashed line corresponding to the fitting using Kondo temperature formula in Appendix \ref{app}.
	}
	\label{fig:Fig8}
\end{figure}

Moving on to Fig. \ref{fig:Fig6}, we further examine the dynamical physical quantities T-matrix and local spin susceptibility. These quantities can be calculated by the convolution of the single particle Green's functions using Eq.~(\ref{Eq:chi_Tmat}). 
The scaling form of $G_f$ and $G_B$ leads to power-law behavior in the T-matrix and susceptibility, with the exponent controlled by $r$. The results validate the statement made in our analytical analysis, particularly the vanishing
anomalous scaling exponent of local spin susceptibility at the critical value of $r$. By utilizing the T-matrix, we can calculate Kondo contribution of resistivity from Eq.~(\ref{Eq:rho}). For $r\leq 0$ case, the T-matrix is either constant or singular for low frequency at zero temperature, resulting in NFL behavior in resistivity, as shown in Fig.~\ref{fig:Fig7}(a). However, the resistivity behaves differently for $r>0$, where exponent equals to $2r$, leading to a decrease of the resistivity when lower the temperature.

It is also interesting to see how impurity entropy $s(T)$ evolution on parameters. We address this using Eq.~(\ref{eq:S}) in Fig.~\ref{fig:Fig8}. Fig.~\ref{fig:Fig8}(a) illustrates the impurity entropy as a function of temperature for different values of $r$. At high temperatures, all curves converge to a value close to the free moment case, which is $\ln 2$. As the temperature decreases, the entropy starts to decline at a characteristic temperature corresponding to $T_K$, and this increases with higher values of $r$ due to the enhanced DOS at the Fermi level. Fig.~\ref{fig:Fig8}(b) depicts the finite temperature phase diagram of the model with a fixed $r=r_c$, as investigated through the impurity entropy. Two distinct phases, distinguished by different impurity entropy values, are separated by a crossover region. The high-temperature free moment phase contracts to zero at zero temperature, and the overscreened phase dominates the entire zero-temperature region, indicating the unstable nature of the LM fixed point. The crossover temperature aligns well with the Kondo temperature obtained analytically in the Appendix \ref{app}.  Fig.~\ref{fig:Fig8}(a) also demonstrates that the impurity entropy at zero temperature is a function of $r$ which decreases as increasing of $r$.

\section{Discussions and conclusions}\label{conclusion}
A natural consideration in the context of the large-$N$ limit is whether the system remains stable when we account for fluctuations in the finite $N$ case. It has been demonstrated that $N=2$ continuous-time quantum Monte Carlo results agree well with the dynamical large-$N$ results\cite{cai2020dynamical}. This agreement can be understood by following the argument presented in Ref. \onlinecite{cox1993spin}. Here, we demonstrate that the saddle-point exponents remain unchanged at the $1/N$ level in our DPLMCK model. In the functional integral formulation, a generic diagram with $L$ loops involving $\Im G_f^{-1}$ must contain $L$ propagators of the $B$-field, $L$ propagators of conduction electrons, and $L-1$ propagators of spinons.  For example, $L=4$ in the diagram of Fig.~\ref{fig:Fig3} (c,d). The most singular part, therefore, behaves as $\delta\Im G_f^{-1}(\omega)\propto |\omega|^{\zeta_f(L)}/N$ where $\zeta_f(L)=L-r L+(\Delta_2-1)L+(L-1)(\Delta_1-1)$. At the saddle point with the relation $\Delta_1+\Delta_2=1+r$, we have $\zeta_f(L)$ fulfills $\zeta_f(L)=1-\Delta_1$. Similarly,  we find $\delta \Im G_B^{-1}\propto |\omega|^{\zeta_B(L)}/N$, where $\zeta_B=1-\Delta_2$. Importantly, this argument holds for both the NFL and pseudogap phases. Thus, fluctuations are irrelevant in altering the exponents governed by saddle points, ensuring the robustness of the pseudogap phase against $1/N$ fluctuations.

In conclusion, we investigated the power-law multichannel Kondo model both analytically and numerically within the large-$N$ limit. At the overscreened fixed point, we observed scaling behaviors for the spinons and $B$ fields, leading to a crossover from the non-Fermi liquid to pseudogap behavior in the spectral properties of spinons by tuning the exponent $r$ or the ratio $\kappa$.
At the critical value of $r_c$ with a fixed $\kappa$, both the resistivity and dynamical local spin susceptibility exhibit power-law behaviors, with $\rho(T)\propto T^{2r_c}$ and $\chi(\omega)\propto \omega$. These vanishing power-law behaviors of resistivity and local susceptibilities deviate from the usual NFL, marginal Fermi liquid, and Landau-Fermi liquid behaviors. Finally, we mention possible applications of our results. In twisted bilayer graphene systems where the DOS exhibits power-law divergence at the magic angle. In these systems, a local moment can be introduced either by substitution or effectively through intrinsic AA-stack configurations of the system\cite{song2022magic,shi2022heavy}. The valley degeneracy of graphene can serve as an independent channel, leading to multichannel physics\cite{sengupta2008tuning}. Our work suggests the potential existence of a pseudogap-type spectrum in spinon excitations in these systems. While the spinon considered in our study is primarily local, a recent study demonstrated that the lattice version of the model, with minimal Heisenberg coupling, spontaneously develops spinon dispersion\cite{ge2022emergent}. In spin liquid system, the itinerant spinons behave like normal electrons, leading to the development of the Kondo screen observed by scanning tunneling spectroscopy\cite{chen2022evidence}. Future studies could explore whether such a pseudogap spectrum of local spinons could persist on the lattice, where spinons is itinerant, giving rise to a pseudogap phase analogous to its counterpart in electrons in the cuprate system. One could also extend the current multichannel model to particle-hole asymmetric case, considering the impact of potential scattering in real systems. Additionally, an extension of the study to the ferromagnetic side could provide insights into the properties of the ferromagnetic fixed point.

\section{Acknowledgments}
X.L.H acknowledges the supports from China Postdoctoral Science Foundation Fellowship (No. 2022M723112). Z.D.Y acknowledges the supports from National Natural Science Foundation of China Grants No. 12204411 and No. 12075205. D.Q.H acknowledges the supports from National Natural Science Foundation of China under Grant No. 12147102.

\appendix
\section{The Kondo temperature for DPLMCK model}\label{app}
We start with following flow equation for multichannel Kondo model with a constant DOS\cite{gan1994on}:
\bea
\frac{\partial \bar{J}}{\partial \ln \omega}=-\bar{J}^{2}+\frac{\kappa}{2} \bar{J}^{3}.
\eea
This model exhibits an intermediate fixed point located at $\bar{J}^{*}=2/\kappa$. The initial value of $\bar{J}$ is $\bar{J}(\omega=D)=\bar{J}_K$. The flow equation can then be reformulated as:
\bea
\frac{\partial \bar{J}}{\partial \ln \omega}=\frac{2}{\kappa}(\bar{J}-\bar{J}^{*})(\frac{\bar{J}}{\bar{J}^{*}})^{2}.
\eea
Let's introduce the parameter $\Delta=2/\kappa$, which corresponds to the slope of the beta function. The solution to the flow equation can be obtained through integration:
\bea
\int_{\bar{J}_K}^{\bar{J}}\frac{d\bar{J}}{\Delta(\bar{J}-\bar{J}^{*})(\frac{\bar{J}}{\bar{J}^{*}})^{2}}=\int_{D}^{\omega}d \ln \omega.
\eea
This integration yields
\bea
\frac{1}{\Delta}[(\frac{\bar{J}^{*}}{\bar{J}}-\ln\frac{\bar{J}}{\bar{J}-\bar{J}^{*}})-(\frac{\bar{J}^{*}}{\bar{J}_K}-\ln\frac{\bar{J}_{K}}{\bar{J}_{K}-\bar{J}^{*}})]= \ln \frac{\omega}{D},
\eea
which can be further expressed as:
\bea
(\bar{J}-\bar{J}^{*})=(\bar{J}_{K}-\bar{J}^{*})(\frac{\omega}{D\bar{J}_{K}^{\frac{1}{\Delta}}e^{-\frac{\bar{J}^{*}}{\Delta \bar{J}_K}}})^{\Delta}\bar{J}e^{-\frac{\bar{J}^{*}}{\bar{J}}}.
\eea
The Kondo temperature can then be defined as
\bea
T_{K}=D\bar{J}_{K}^{\frac{1}{\Delta}}e^{-\frac{\bar{J}^{*}}{\Delta \bar{J}_K}}=D\bar{J}_{K}^{\frac{\kappa}{2}}e^{-\frac{1}{\bar{J}_K}}.
\label{eq:Kondo1}
\eea

Now, let's consider the power-law DOS case. For the single-channel Kondo model, the flow equation is\cite{mitchell2013quantum}
\bea
\frac{\partial \bar{J}}{\partial \ln \omega}=-r\bar{J}-\bar{J}^{2}.
\eea
The strong coupling fixed point is at infinity. Consequently, upon integration, the equation becomes:
\bea
\int_{\bar{J}_{K}}^{\gg 1}\frac{d\bar{J}}{-r \bar{J}-\bar{J}^{2}}=\int_{D}^{\omega^{*}} d\ln \omega.
\eea
Here, $\omega^{*}$ is the energy regime where $\bar{J}\gg 1$, indicating the breakdown of perturbation theory. Thus $T_K=\omega^{*}$. Integration yields
\bea
\frac{\ln\frac{\bar{J}_K}{\bar{J}_K+r}}{r}=\ln(\frac{T_K}{D}),
\eea
leading to
\bea
T_K=D(1+\frac{r}{\bar{J}_K})^{-1/r}.
\label{eq:Kondo2}
\eea

Based on two special cases mentioned above, we are now ready to derive the Kondo temperature for the flow equation under consideration in this paper:
\bea
\frac{\partial \bar{J}}{\partial \ln \omega}=-(r \bar{J}+\bar{J}^{2}-\kappa \bar{J}^{3}).
\eea
The fixed point is given by $\bar{J}^{*}=\dfrac{1+\sqrt{1+4 \kappa r}}{2\kappa}$. The flow equation can be expressed in the following form:
\bea
\frac{\partial \bar{J}}{\partial \ln \omega}=(\bar{J}-\bar{J}^{*})\bar{J}(\frac{r}{\bar{J}^{*}}+\kappa \bar{J}).
\eea
Solving this equation through integration yields:
\begin{widetext}
\bea
\frac{1}{r}\left[-\ln \bar{J}+\ln(\bar{J}-\bar{J}^{*})^{\frac{r}{(r+\bar{J}^{*2}\kappa)}}+\ln(r+\kappa \bar{J}^{*}\bar{J})^{\frac{\bar{J}^{*2}\kappa}{r+\bar{J}^{*2}\kappa}}\right]\bigg|_{\bar{J}_K}^{\bar{J}}=\ln\frac{\omega}{D}.
\eea
Expressed in the following form:
\bea
\bar{J}-\bar{J}^{*}=(\bar{J}_K-\bar{J}^{*})(\frac{\omega}{DJ_{K}^{1/r}(r+\kappa \bar{J}^{*}\bar{J}_K)^{-\frac{\bar{J}^{*2}\kappa}{r(r+\bar{J}^{*2}\kappa)}}})^{r+\bar{J}^{*2}\kappa}\left[\bar{J}^{1/r}(r+\kappa \bar{J}^{*}\bar{J})^{-\frac{\bar{J}^{*2}\kappa}{r(r+\bar{J}^{*2}\kappa})}\right]^{r+\bar{J}^{*2}\kappa}
\eea
\end{widetext}
Therefore, the Kondo temperature is
\bea\label{eq:Tk}
T_K=&D\bar{J}_{K}^{1/r}(r+\kappa \bar{J}^{*}\bar{J}_K)^{-\frac{\bar{J}^{*2}\kappa}{r(r+\bar{J}^{*2}\kappa)}},
\label{eq:Kondo3}
\eea
it is straightforward to verify that Eq.~(\ref{eq:Kondo3}) reduces to Eq.~(\ref{eq:Kondo2}), or Eq.~(\ref{eq:Kondo1}) when $\kappa\rightarrow 0$, or $r\rightarrow 0$, respectively.

\bibliographystyle{apsrev4-1}

%


\end{document}